\documentclass[11pt,a4paper]{article}

\usepackage{fullpage}
\usepackage{indentfirst}
\usepackage[utf8x]{inputenc}
\usepackage[T1]{fontenc}
\usepackage{hyperref}
\usepackage[english]{babel}
\usepackage[fleqn]{amsmath}
\usepackage{amsfonts}
\usepackage{amssymb}
\usepackage{amscd}
\def \be{\begin{equation}}
\def \ee{\end{equation}}

\def \d{{\rm d}}
\author{V. Aldaya$^{1}$, F. Coss\'{\i}o$^{1}$, J. Guerrero$^{1,2}$ and F.F.
L\'opez-Ruiz$^{1}$}

\title{The quantum Arnold transformation}

\date{\begin{center}
\begin{small}$^1$ Instituto de Astrof\'{\i}sica de Andaluc\'{\i}a, CSIC, 
\end{small}\\	
\begin{small}
  Apartado Postal 3004, 18080 Granada, Spain
\end{small}\\
\begin{small}$^2$Departamento de Matem\'atica Aplicada, Universidad de
Murcia, \end{small}\\
\begin{small}Campus de Espinardo, 30100 Murcia, Spain.\end{small}\\
\begin{small}valdaya@iaa.es\ fcossiop@gmail.es \ juguerre@um.es\ flopez@iaa.es
\end{small}\\                                                                   
\end{center}
}
\begin{document}

\maketitle
%

%\tableofcontents

\begin{abstract}
By a quantum version of the Arnold transformation of classical mechanics, all
quantum dynamical systems whose classical equations of motion are
non-homogeneous linear second-order ordinary differential equations, including
systems with friction linear in velocity, can be related to the quantum 
free-particle dynamical system. This transformation provides a basic
(Heisenberg-Weyl) algebra of quantum operators, along with well-defined
Hermitian operators which can be chosen as evolution-like observables and
complete the entire Schr\"odinger algebra. It also proves to be very helpful in
performing certain computations quickly, to obtain, for example, wave functions
and closed analytic expressions for time-evolution operators.

\end{abstract}

% Keywords: Arnold transformation, non-homogeneous linear second order 
% differential equations, damped harmonic oscillator, symmetry groups, 

%\tableofcontents{}

\section{Introduction}

The description of the quantum damped harmonic oscillator by
the Caldirola-Kanai model \cite{Caldirola,Kanai}, which involves a
time-dependent Hamiltonian, has attracted the attention of many authors, as
could be considered one of the simplest and paradigmatic examples of dissipative
system. In particular, its analysis from the symmetry point of view has
proved to be very fruitful. In a purely classical context, the symmetries of the
equation of the damped harmonic oscillator with time-dependent parameters were
found in \cite{Martini}. In \cite{Cervero-C} Cerver\'o and Villarroel found,
for the damped harmonic oscillator, finite-dimensional point symmetry groups for
the corresponding Lagrangian (the un-extended Schr\"odinger group
\cite{schrogroup}) and the equations of motion ($SL(3,\mathbb{R})$)
respectively, and an infinite contact one for the set of trajectories of the
classical equation. They singled out a ``non-conventional'' Hamiltonian from
those generators of the symmetry, recovering some results from
\cite{Hioe-Yuen,Manko}.
% Then, they concluded that the damped harmonic oscillator should not be claimed
% to be dissipative at all at the quantum level, as this ``non-conventional''
% Hamiltonian is conserved, and should be related to an oscillator with variable
% frequency.

Some flaws has been associated with the quantum description of the damped 
harmonic oscillator according to Caldirola-Kanai equation. For instance, 
it is claimed that uncertainty  relations are not preserved under time evolution 
and could eventually be violated \cite{Brittin,Razavy_libro}, although this 
inconsistency seems to be associated  with a confussion between canonical 
momentum and ``physical'' momentum  \cite{Schuch}.

There exists another interesting approach to the study of the classical damped
harmonic oscillator, based on the observation that its classical equation of
motion is a special case of the set of linear second-order ordinary differential
equations (LSODE for short). In Classical Mechanics the family of solutions of a
second-order differential equation corresponding to the motion of a given
physical problem is sometimes related to that of a simpler system, considered as
a toy model, in order to import from it simple general properties which could be
hidden in the real problem. Both physical systems should share global properties
of the solution manifold, such as topology and symplectic structure. The
paradigmatic example is the transformation described by Arnold in \cite{Arnold},
which brings any LSODE to the simplest form of the free Galilean particle
equation.
This transformation turns out to be extremely useful. In particular, it is
possible to obtain the symmetry group of a particular instance of LSODE
\cite{Martini}, in which the symmetries of the action of the corresponding
system can be found as a subgroup \cite{Lutzky}.

Therefore, it seems natural to try to generalize the Arnold transformation to
the quantum level, to be denoted as Quantum Arnold Transformation (QAT), as much
insight can be gained in the study of any system classically described by a
LSODE and, in particular, the parametric oscillator or some of the
systems which present dissipation.

Several partial generalizations can be found in the literature. For example, in
\cite{Takagi}, Takagi is able to provide a transformation which relates the
Schr\"odinger equation of the harmonic oscillator to that of the free
particle, and applies it to simplify the computation of the propagator by making
use of the free one. \cite{Huang-Wu} contains a slightly more general version
(see formula (33) therein). \cite{Kanasugi} went a bit further considering the
damped harmonic oscillator with constant parameters. None of them mentions the
classical Arnold transformation, but it \textit{is} underlying their reasoning. 

Implicitly, a generalization of the Arnold transformation was also contained in
\cite{Manko}, the classical version not being referred once more.
It will be shown that some of their results formally converge with ours (see
Section \ref{exampledamposc}), although they put emphasis on another aspects of
the problem, such as the analysis of unitarity and energy loss.
Mostafazadeh \cite{Mostafa1} also pursued the idea of
``connecting'' different quantum physical systems by means of time-dependent
unitary transformations, even representing arbitrary
time-dependent diffeomorphisms \cite{Mostafa2}. His approach is rather general,
but does not fully take advantage of the possibility of connecting with the free
particle system and importing its symmetries.

The relevance of generalizing the Arnold transformation is that it
can be used to export properties from the free Schr\"odinger equation to that of
the system corresponding to the given classical equation: the complete set of
symmetries of the quantum free particle, the Schr\"odinger group
\cite{schrogroup,key-1}, can be realized on the system under study, providing as many
conserved quantities as in the free particle. In particular, the generators
corresponding to the free position and momentum prove to be good basic quantum
operators, constituting a quantization. In addition, the transformation turns
out to be extremely useful to compute objects that would otherwise need
laborious calculations, such as wave functions, the quantum propagator or the
evolution operator. 

We shall see that time translations in the non-free system do not belong,
in general, to the imported Schr\"odinger group. This is to be expected, as the
energy in this system is not conserved when the classical equation of motion
includes a friction term or  a time-varying frequency.
A deeper analysis of this fact can be found in 
\cite{batenuestro}.

To obtain the QAT, we begin in Section \ref{clasarnold} by observing that a
generalized version of the classical equation of the damped harmonic oscillator,
in which the constant coefficients are promoted to be time-dependent, can be
transformed by a local diffeomorphism of position and time, the Arnold
transformation, into the equation of motion of the free particle. In Section
\ref{quanarnold} we generalize this interesting feature from the classical to
the quantum theory. The Caldirola-Kanai equation %\cite{Caldirola,Kanai} 
for the damped harmonic oscillator is then a particular case of the general one.
In Section \ref{examples} we illustrate the use of QAT in a couple of simple
examples presenting damping, the damped particle and the damped harmonic
oscillator.

\section{Classical Arnold transformation}
\label{clasarnold}

Mathematically speaking, the classical Arnold transformation \cite{Arnold}
converts any linear second-order ordinary differential equation (LSODE) into
the free Galilean particle equation, that is, $\ddot{\kappa}=0$ in 1+1
dimensions (we shall limit ourselves to this situation).

From the physical point of view, the Arnold transformation relates
the trajectories $x(t)$, with initial conditions $x_0$ and $p_0\sim \dot x_0$,
solutions of the LSODE, to those trajectories $\kappa (\tau)$ solutions of the
free equation with initial conditions $\kappa_0$ and $\pi_0\sim \dot \kappa_0$.
Either $(x_0,p_0)$ or $(\kappa_0,\pi_0)$ parametrize the common solution
manifold $\mathcal M$, and we shall adopt the unified notation $(K,P)$. On this
manifold, each physical system is characterized by the corresponding Hamiltonian
as a function of $K$ and $P$. The inverse of the corresponding Hamilton-Jacobi
transformation then recovers the trajectories $(x(t),\dot x(t))$ or $(\kappa
(\tau),\dot\kappa(\tau))$ out of the $K,P$ variables. 

Following a similar notation to that in \cite{Martini}, we give an overview
of the Arnold transformation \cite{Arnold}. Firstly, let us recall that, given
an arbitrary, non-homogeneous LSODE
\begin{equation}
\ddot{x}+\dot{f}\dot{x}+\omega^{2}x=\Lambda,\label{eq:ecuacion no homogenea}
\end{equation}
where $\dot{x} = \frac{d x}{d t}$ and so on, and $f$, $\omega$ and $\Lambda$ are
arbitrary functions of time $t$, we can apply the transformation
\begin{equation}
\begin{cases}
t\longrightarrow t\\
x\longrightarrow x+u_p\end{cases},\label{eq:difeomorfismo no homogeneo}
\end{equation}
$u_p$ being a particular solution of (\ref{eq:ecuacion no homogenea}). We
find that the differential equation above is transformed into
\begin{equation}
\ddot{x}+\dot{f}\dot{x}+\omega^{2}x=0,\label{eq:ecuacion lineal homogenea}
\end{equation}
i. e., every non-homogeneous problem is equivalent to a homogeneous one. 

The homogeneous Arnold transformation, is a local
diffeomorphism which maps the free particle equation of motion into
(\ref{eq:ecuacion lineal homogenea}):
\begin{equation}
\begin{cases}
\tau=\frac{u_{1}(t)}{u_{2}(t)}\\
\kappa=\frac{x}{u_{2}(t)}
\end{cases},
\qquad
\ddot \kappa =0 \quad \longleftrightarrow \quad
\ddot{x}+\dot{f}\dot{x}
+\omega^{2}x=0 \, ,
\label{eq:difeomorfismo}
\end{equation}
where $u_{1}(t)$ and $u_{2}(t)$ are independent solutions  of (\ref{eq:ecuacion
lineal homogenea}). Applying the inverse diffeomorphism to the classical
dynamical system (\ref{eq:ecuacion lineal homogenea}), we can transform this
equation into the free one. 

If we include external forces the transformation \eqref{eq:difeomorfismo}
turns into the general Arnold transformation, that we shall call simply $A$:
\begin{equation}
\begin{cases}
\tau=\frac{u_{1}(t)}{u_{2}(t)}\\
\kappa=\frac{x - u_p(t)}{u_{2}(t)}
\end{cases},
\qquad
\ddot \kappa =0 \quad \longleftrightarrow \quad
\ddot{x}+\dot{f}\dot{x} +\omega^{2}x=\Lambda \, .
\label{eq:difeomorfismo no homogeneo completo}
\end{equation}

This transformation could be understood as
passing to coordinates analogous to co-moving spacial coordinate and proper
time used in General Relativity, so that the system becomes ``free'', at least
locally.

Indeed this transformation is of \textit{local} nature in time, in the sense
that it is only valid for an open interval in time $t$. In fact, it can be shown
that the equivalence in \eqref{eq:difeomorfismo}  and \eqref{eq:difeomorfismo no
homogeneo completo} is true up to a factor 
$\frac{u_2^2}{e^{-2f}}$, so that it holds in the interval where
$u_2$ does not vanish. 
This means that the transformation does not take the Euler-Lagrange
operator associated with the LSODE itself to that of the free system. For this
reason it can not be claimed that both physical systems are actually
equivalent. 

However, Arnold transformation can help to understand the physical system under
study. In particular, as pointed in \cite{Martini}, it is possible to identify
the set of contact symmetries for (\ref{eq:ecuacion lineal homogenea}), and
this way to arrive at the results found in \cite{Cervero-C}, which show the
sets of symmetries for either the equations or the action from which such
equations can be derived. 

It should be noted that, due to the general character of the transformation, we
could miss the physical identity of position and time when performing such a
transformation. But it will be possible to choose appropriate
specific solutions $u_1$, $u_2$ and $u_p$ with suitable initial conditions so
that the identity of the variables is maintained\footnote{For his restricted
version of the quantum Arnold transformation, Takagi, in \cite{Takagi},
suggested that the transformation could be specified by different solutions at
different times, in principle avoiding the restriction of locality.
However, care must be taken when making use of this freedom to prevent conflict
with preserving the identity of the variables when it is desirable.}. We shall
make use of this possibility in going to the quantum version of this
transformation.

Finally, having in mind the particular case of dissipative systems, we would
like to remark that certain issues of the treatment of these systems are
already apparent in the classical domain. For instance, time evolution is not
a symplectomorphism, nor preserve the volume of the phase space. Obviously, the
Hamiltonian function is not a Noether invariant. This becomes especially
manifest and annoying when formulating the quantum theory.

\section{The Quantum Arnold transformation}
\label{quanarnold}

As already mentioned in the Introduction, several partial versions of the QAT
can be found in the literature. Here we give a generalization that contains, as
particular cases, those found in \cite{Takagi,Huang-Wu,Kanasugi,Manko}.

In bringing Arnold's technique to the quantum world we must be aware, obviously,
of the different philosophy of the quantum description and different nature of
the equation of motion. The objects and structures that define a quantum system,
namely the Hilbert space, the basic observables, the Hamiltonian operator, and
the Schr\"odinger equation must be specified in a way that we are able to
identify the same objects at both sides of the transformation. 

To this end, it is important to focus, in the free system, only on those
operators corresponding to constants of motion, Noether invariants associated
with its symmetry, that is, the Schr\"odinger group (which contains the
centrally-extended Galilei group as a subgroup, containing in turn the
Heisenberg-Weyl group of translations and non-relativistic boosts). This implies
to fix the basic operators (that is to say, quantum operators which realize a
unitary and irreducible representation of the common classical Poisson
(Heisenberg-Weyl) algebra $\{K,P\}=1$) so that they respect the Schr\"odinger
equation, then having constant expectation values and being generators of the
basic symmetry\footnote{By ``basic symmetry'' we understand, in general, those
symmetries whose associated Noether invariants are enough to parametrize the
classical solution manifold.} (the Heisenberg-Weyl sub-group of the
centrally-extended Schr\"odinger group).

Those basic operators, in this form, are in principle the candidates to be
related by a quantum version of the Arnold transformation, so that we shall
have the situation as follows: On the one hand, a common Hilbert space $\mathcal
H$ of wave functions $\Psi(K)$
($L^2(\mathbb{R})$), which plays the role of initial values for both the
solutions $\phi(x,t) \in \mathcal H_t$ of the Schr\"odinger equation relative
to the quantum version of our original LSODE and those wave functions,
$\varphi(\kappa,\tau) \in \mathcal H^G_\tau$, solutions of the free
Schr\"odinger equation.
And, on the other hand, the  quantum Arnold transformation $\hat A$ relating
Schr\"odinger equations and basic operators. As a crucial consequence, we
shall obtain a realization of the free symmetry on the quantum, non-free
system.
The following diagram can help to have a picture of the setup:
\begin{equation}
\begin{CD} 
\mathcal H^G_\tau @<\hat A<< \mathcal H_t\\ 
@A{\hat U_G(\tau)}AA @AA\hat U(t)A\\ 
\mathcal H^G_0 \equiv \mathcal H @>>\hat{1}> \mathcal H \equiv \mathcal H_0 
\end{CD}
\label{diagram}
\end{equation}
$\mathcal H^G_\tau$ (resp. $\mathcal H_t$) is the Hilbert space of
solutions of the free or Galilean (resp. non-free or corresponding to the LSODE)
Schr\"odinger equation, $\hat U_G(\tau)$ (resp. $\hat U(t)$) is the free (resp.
non-free) evolution operator and $\hat 1$ is the identity operator. 
Here, the Hilbert space $\mathcal H$ may be considered as the quantum analogue
of the classical solution manifold, $\mathcal M$, usually thought of as space of
(classical) initial conditions. On $\mathcal H$ one must be able to measure all
possibles physical observables in much the same way classical observables are
characterized as real functions on $\mathcal M$, that is, functions whose
arguments are constants along classical trajectories (functions of Noether
invariants).

The Hamiltonian of the non-free system, not being conserved in general, will not
be related to any operator from the free particle. This is to be expected, since
it is not a conserved quantity under the evolution of the physical system,
neither at classical nor at the quantum level (in the sense that it does not
have constant
expectation values). It is important to remark that this implies that it is not
possible to formulate a time-independent Schr\"odinger equation.

More specifically, by extending properly the Arnold transformation (or the
inverse) to the quantum case, we shall relate the space $\mathcal H^G_\tau$ of
solutions of the free Schr\"odinger equation 
\begin{equation}
i\hbar\frac{\partial \varphi}{\partial \tau}=-\frac{\hbar^{2}}{2m}
\frac{\partial^{2} \varphi}{\partial \kappa^{2}}\, ,
\label{eq:Schrodinger libre}
\end{equation}
with corresponding classical equation
\begin{equation}
\ddot{\kappa}=0,%\quad \dot \kappa\equiv\frac{d \kappa}{d \tau},
\label{eq:ecuacion libre}
\end{equation}
to that space $\mathcal H_t$ where the quantum theory of the generic
LSODE
\begin{equation}
\ddot{x}+\dot{f}\dot{x}+\omega^{2}x= \Lambda,
\label{eq:ecuacion de movimiento general}
\end{equation}
is realized, the quantities $f$, $\omega$ and $\Lambda$ being, in general,
time-dependent.

The classical equation (\ref{eq:ecuacion de movimiento general}) can be derived
from a variational principle. We shall consider the Lagrangian
\begin{equation}
  L= e^{f}\Bigl(\frac{1}{2} m {\dot x}^2 -
\frac{1}{2} m \omega^{2}x^{2} + m \Lambda x \Bigr)
\label{eq:lagrangiano general}
\end{equation}
as our starting point. The Schr\"odinger equation can be derived from the
corresponding classical Hamiltonian function
\begin{equation}
H=\frac{p^{2}}{2m}e^{-f}+\bigl(\frac{1}{2}m\omega^{2}x^{2} -m\Lambda x
\bigr)e^{f}\,,
\label{eq:hamiltoniano general}
\end{equation}
according to the standard canonical prescriptions, leading to
\begin{equation}
i\hbar\frac{\partial \phi}{\partial t}=-\frac{\hbar^{2}}{2m}e^{-f}
\frac{\partial^{2} \phi}{\partial x^{2}}+\bigl(\frac{1}{2}m\omega^{2}x^{2}
- m\Lambda x
\bigr)e^{f}\phi .
\label{eq:Schrodinger general}
\end{equation}
For $f$ linear in time, constant $\omega$ and $\Lambda = 0$
this equation is commonly known as Caldirola-Kanai equation for the damped
harmonic oscillator \cite{Caldirola,Kanai}.

Even though both spaces of solutions of (\ref{eq:Schrodinger libre}) and
(\ref{eq:Schrodinger general}), $\mathcal H^G_\tau$ and
$\mathcal H_t$ respectively, will be related, and the basic quantum
operators associated with the classical functions $K,P$ realized as well-defined
operators both on $\mathcal H^G_\tau$ and $\mathcal H_t$, we cannot still
assure that both physical systems are actually equivalent. In fact, the
evolution operator in time $t$, $i\hbar\frac{\partial}{\partial t}$, does
definitely not leave invariant the space of the solutions of
(\ref{eq:Schrodinger general}) in general, nor comes down to the space $\mathcal
H$, which means that it cannot be realized as an operator  function of
$\hat{K},\hat{P}$ (in sharp contrast to $i\hbar\frac{\partial}{\partial \tau}$,
which is $\sim \hat{P}^2$). We shall achieve the construction of well-defined
Hermitian evolution-like generators imported from the free system via the
inverse quantum Arnold transformation, but their eigenvalues, conserved indeed,
do not correspond to the standard energy\footnote{The construction of a properly
defined, Hermitian generator in standard time will take much effort requiring a
sound analysis of the symmetry problem in damped systems. This has been done in
\cite{batenuestro}.}.
These operators close the Schr\"odinger algebra with the basic operators. 

Implicitly, this trick of considering operators different from the
Hamiltonian to provide quantum numbers and obtain solutions of the
Schr\"odinger equation as their eigenfunctions has been used
extensively. 
For example, the operator found in \cite{Manko} to label the
energy-loss states, which coincides with the quantum operator $H^{*}$
corresponding to $G_5$ in
\cite{Cervero-C,Cervero-Q}, turns out to be a generator of the  $SL(2,\mathbb R)$
subgroup of the Schr\"odinger group.
We give here explicitly the frame in which this can be done: operators
from the Schr\"odinger group can be chosen to play this role upon convenience. 

\bigskip

For the sake of simplicity, let us focus on the case with no external forces
$\Lambda = 0$. The formulas corresponding to $\Lambda \neq 0$ are given
in Appendix \ref{appa}. 

The generalization of the classical Arnold transformation is
obtained by completing (\ref{eq:difeomorfismo}) with a 
change of the wave function.
Explicitly, the quantum Arnold transformation, valid for every physical system
with classical equation of the homogeneous LSODE type, is given by the (local)
diffeomorphism:
\begin{equation}
\begin{cases}
\tau=\frac{u_{1}(t)}{u_{2}(t)}\\
\kappa=\frac{x}{u_{2}(t)} \\
\varphi =\phi \sqrt{u_{2}(t)}\,e^{-\frac{i}{2}\frac{m}{\hbar}
\frac{1}{W(t)}\frac{\dot{u}_{2}(t)}{u_{2}(t)}
{x}^{2}}\; ,
\end{cases}
\label{eq:quantum Arnold transformation}
\end{equation}
where $u_1$ and $u_2$ satisfy again the classical equation of motion in $(x,t)$,
$\dot{u}_{1} = \frac{d u_{1}}{d t},\dot{u}_{2} = \frac{d u_{2}}{d t}$ and
$W(t) \equiv \dot{u}_{1} u_{2} - u_{1} \dot{u}_{2}=e^{-f}$. It is
straightforward to check that by this transformation the Schr\"odinger equation
of the free particle is transformed into (\ref{eq:Schrodinger general}) (with
$\Lambda = 0$) up to a
multiplicative factor which depends on the particular choice of the classical
solutions $u_1$ and $u_2$ (partial derivatives must be changed by the
classical part of the transformation while wave functions are shifted by the
quantum part).

Now, we can impose on $u_1$ and $u_2$ the condition that they preserve the
identity of $\tau$ and $\kappa$, i.e., that $(\kappa,\tau)$ coincide with $(x,
t)$ at an initial point $t_0$, arbitrarily taken to be $t_0=0$:
\begin{equation}
 u_1(0)=0, \; u_2(0)=1, \qquad \dot{u}_1(0)=1, \; \dot{u}_2(0)=0 \,. 
\label{eq:condiciones} 
\end{equation}
This fixes a unique form of the diffeomorphism for a given ``target'' physical
system. However, the quantum Arnold transformation would still be valid
if solutions $u_1$ and $u_2$ do not satisfy \eqref{eq:condiciones}. The price
to be paid would then be that the relation in the lower part of the diagram
\eqref{diagram} above would no longer be the identity and basic position and
momentum operators would then be mixed (see the end of this Section).

Formally, while the classical Arnold transformation in this case is:
\begin{equation}
\begin{split}
  A:& \; \; \mathbb R \times T \longrightarrow \mathbb R \times T' \\
    & \quad (x,t)  \longmapsto \; (\kappa,\tau) =
        A\bigl((x,t)\bigr) = (\tfrac{x}{u_2},\tfrac{u_1}{u_2})\,,
\end{split}
\end{equation}
where $T$ and $T'$ are open intervals of the real line containing $t=0$
and $\tau=0$, respectively, QAT can be written:
\begin{equation}
\begin{split}
  \hat A:& \quad \mathcal H_t \;\; \longrightarrow \quad \mathcal H^G_\tau \\
    & \phi(x,t)  \longmapsto \; \varphi(\kappa,\tau) = 
        \hat A \bigl( \phi(x,t) \bigr) = 
       A^* \bigl( \sqrt{u_{2}(t)}\,e^{-\frac{i}{2}\frac{m}{\hbar}
                 \frac{1}{W(t)}\frac{\dot{u}_{2}(t)}{u_{2}(t)}
                                                 {x}^{2}} \phi(x,t) \bigr) \,,
\end{split}
\end{equation}
where $A^*$ denotes the pullback operation corresponding to $A$. 

As already remarked, the basic symmetries of the free system are inherited by
the LSODE-type system, as we are now able to transform the infinitesimal
generators of translations (the Galilean momentum operator $\hat \pi$,
corresponding to the classical conserved quantity `momentum') and
non-relativistic boosts (the position operator $\hat \kappa$, corresponding to
the classical conserved quantity `initial position'). They are, explicitly,
\begin{align}
\hat \pi &= -i \hbar \frac{\partial}{\partial \kappa} 
\label{eq:operata momento libre}
\\
\hat \kappa &= \kappa + \frac{i \hbar}{m} \tau
\frac{\partial}{\partial \kappa}\;, 
\label{eq:operata posicion libre}
\end{align}
that is, those basic, canonically commuting operators with constant expectation
values, that respect the solutions of the free Schr\"odinger
equation, have constant matrix elements (and constant expectation values in
particular) and fall down to well defined, time-independent operators in the
Hilbert space of the free particle $L^2(\mathbb R)$.

In general, these properties are satisfied whenever an operator $\hat O(t)$
can be written as  
\begin{equation}
 \hat O(t)= \hat U(t,t_0) \; \hat O \; \hat U^\dagger(t,t_0)\, ,
 \label{eq:operadorsubido}
\end{equation}
where $\hat O$ is $\hat O(t_0)$ and $\hat U(t,t_0)$ is the evolution
operator satisfying the Schr\"odinger equation\footnote{Note
that $\hat O(t)$ is \textit{not} the usual Heisenberg picture version 
$\hat O_H(t) = \hat U(t,t_0)^\dagger \; \hat O \; \hat U(t,t_0)$
of its associated operator in Schr\"odinger picture $\hat O$, although their
relation is very simple when the Hamiltonian is time-independent.}. If the
Hamiltonian is time-independent, as in the free particle case, time-evolution is
a one-parameter group and then $\hat U(\tau,\tau_0)= \hat U(\tau-\tau_0)$.

Defining a generic Schr\"odinger equation operator, $\hat S \equiv i \hbar
\frac{\partial}{\partial t}-\hat H$, taking $t_0=0$ ($\hat U(t) \equiv \hat
U(t,0)$) and reminding that $\hat U(t,t_0)^\dagger\hat U(t,t_0)=1$, 
it is clear that, for operators of the form (\ref{eq:operadorsubido}):
\begin{gather}
 \hat S \,\hat O (t) \,|\psi (t)\rangle = 
 \hat S \,\hat U(t)\, \hat O\, |\psi \rangle \equiv 
 \hat S \,\hat U(t)\, |\psi' \rangle = 
 \hat S\, |\psi'(t) \rangle = 0
 \\
 \frac{\partial}{\partial t}\langle \chi(t)|\hat O (t)|\psi (t)\rangle = 0
 \\
 \frac{\partial}{\partial t}\left(\hat U(t)^\dagger \hat O(t) \hat U(t)
  \right)= 0
 \\
 \frac{\d}{\d t} \hat O(t) \equiv \frac{\partial}{\partial t} \hat O(t) 
   + \frac{i}{\hbar}[\hat H(t),\hat O(t)] = 0\,
\end{gather}
(where $|\psi (t)\rangle \equiv \hat U(t) |\psi \rangle$), stating that those
operators respect solutions, have constant matrix elements, fall down to define
time-independent operators on the Hilbert space and are integers of the motion,
respectively.
(\ref{eq:operata momento libre}) and (\ref{eq:operata posicion libre}) can be
``de-evolved'', so that  
\begin{equation}
  \hat U(\tau)^\dagger \,\hat \pi \, \hat U(\tau)\,=\,
      -i\hbar \frac{\partial}{\partial \kappa}\,, \qquad
  \hat U(\tau)^\dagger \, \hat \kappa \, \hat U(\tau)\, =\,\kappa \,,
\end{equation}
thus showing the properties above for the free particle.

What we are doing is to focus on these integrals of motion, $\hat \pi$ and
$\hat \kappa$, so  that the new operators position $\hat X$ and momentum $\hat
P$ acting on $\mathcal H_t$ are also integrals of motion in the non-free system.
These will be the generators of the basic symmetry in the non-free system.
Dodonov and Man'ko \cite{Manko} obtained these operators in particular cases by
direct calculation, imposing them to commute with the Schr\"odinger equation
$\hat S$. The difference is that, having related this system
with that of the free particle, now it is clear how far one can go: the
Schr\"odinger group and its enveloping algebra. Even more, the approach followed
in \cite{Manko} is only
able to provide the basic operators (corresponding to linear functions in the
classical solution manifold), since the condition of commuting with 
$\hat S$, $[\hat S,\hat O(t)]=0$, is more restrictive than that considered here
of respecting solutions, which is equivalent to 
$[\hat S,\hat O(t)]\sim \hat S$.  

\bigskip

Let us apply the QAT (\ref{eq:quantum Arnold
transformation}) to (\ref{eq:operata momento libre}) and (\ref{eq:operata
posicion libre}). For a given operator $\hat \pi$ acting on $\mathcal H^G_\tau$
there is a corresponding operator $\hat P = \hat A^{-1} \hat \pi \hat A$ on
$\mathcal H_t$. 
The action on functions $\phi(x,t)$ can then be obtained as follows: 
\begin{equation}
  \begin{split}
 \hat P \phi(x,t) &= 
   \hat A^{-1} \hat \pi \hat A \phi(x,t) = 
   \hat A^{-1} \hat \pi A^{*}
       \bigl(   \sqrt{u_{2}}\,e^{-\frac{i}{2}\frac{m}{\hbar}
                 \frac{1}{W}\frac{\dot{u}_{2}}{u_{2}} {x}^{2}} \phi(x,t)
       \bigr) 
= \\ 
&= \frac{1}{\sqrt{u_{2}}} e^{+\frac{i}{2}\frac{m}{\hbar}
                 \frac{1}{W}\frac{\dot{u}_{2}}{u_{2}} {x}^{2}}
   A^{*-1} \Bigl( - i \hbar \frac{\partial}{\partial \kappa} 
      A^{*}  \bigl(   \sqrt{u_{2}}\,e^{-\frac{i}{2}\frac{m}{\hbar}
                 \frac{1}{W}\frac{\dot{u}_{2}}{u_{2}} {x}^{2}} \phi(x,t)
              \bigr)
            \Bigr)
= \\ 
&=   \frac{1}{\sqrt{u_{2}}} e^{+\frac{i}{2}\frac{m}{\hbar}
                 \frac{1}{W}\frac{\dot{u}_{2}}{u_{2}} {x}^{2}}
       \Bigl( - i \hbar u_2 \frac{\partial}{\partial x} 
          \bigl(   \sqrt{u_{2}}\,e^{-\frac{i}{2}\frac{m}{\hbar}
                       \frac{1}{W}\frac{\dot{u}_{2}}{u_{2}} {x}^{2}} \phi(x,t)
           \bigr)
        \Bigr) 
= \\ 
&=      \bigl( -i \hbar u_2 \frac{\partial}{\partial x} 
             - m \frac{\dot u_2}{W} x 
         \bigr) \phi(x,t) \,.
\end{split}
\end{equation}
We can perform the same computation for the position operator and then we have:
\begin{align}
\hat P &= -i \hbar u_2 \frac{\partial}{\partial x} - m x \frac{\dot{u}_2}{W}
\label{eq:operata momento LSODE}
\\
\hat X &= \frac{\dot{u}_1}{W}x + \frac{i\hbar}{m} u_1
\frac{\partial}{\partial x}\, ,
\label{eq:operata posicion LSODE}
\end{align}
thus providing the generators of the realization of the (centrally-extended)
Heisenberg-Weyl symmetry on the physical system corresponding to a general
LSODE.

The properties of the operators (\ref{eq:operata momento LSODE}) and
(\ref{eq:operata posicion LSODE}), i.e.\ preserving solutions of
(\ref{eq:Schrodinger general}), having constant matrix
elements, falling to the Hilbert space, are ensured by the properties of
(\ref{eq:operata momento libre}) and (\ref{eq:operata posicion libre}) before
the transformation, and will be explicitly checked for some particular cases in
next Section. It will also become clear that the identity of both operators is
preserved after the transformation. 

Apart from $\hat P$ and $\hat X$, we can compute $\hat P^2$, $\hat X^2$ and
$\hat{XP}\equiv \frac{1}{2}(\hat X \hat P + \hat P \hat X)$: 
\begin{align}
\hat P ^2 &= -\hbar^2 u_2^2  \frac{\partial^2}{\partial x^2} + 
i \hbar \frac{2 m u_2 \dot u_2}{W} \,x\, \frac{\partial}{\partial x}
 + m^2 \frac{\dot u_2^2}{W^2} \, x^2 + i \hbar \frac{m u_2 \dot u_2}{W}
\label{eq:cuadraticos1}
\\
\hat X ^2 &= \frac{\dot u_1^2}{W^2} \, x^2   
+ i \hbar \frac{2 u_1 \dot u_1}{m W} \,x\,  \frac{\partial}{\partial x}
-\hbar^2 \frac{u_1^2}{m^2}  \frac{\partial^2}{\partial x^2} 
+ i \hbar \frac{u_1 \dot u_1}{m W}
\label{eq:cuadraticos2}
\\
\hat{XP} & =
\frac{\hbar^2}{m} u_1 u_2  \frac{\partial^2}{\partial x^2} 
- i \hbar\frac{\dot u_1 u_2 + u_1 \dot u_2}{W} \,x\, \frac{\partial}{\partial x}
 - m \frac{\dot u_1 \dot u_2 }{W^2} \, x^2
- i  \hbar \frac{\dot u_1 u_2 + u_1 \dot u_2}{2 W}
\,.
\label{eq:cuadraticos3}
\end{align}
Their first-order versions, valid on solutions of 
\eqref{eq:Schrodinger general}, are:
\begin{align}
\hat P ^2 &= i\hbar \frac{2 m u_2^2}{W}  \frac{\partial}{\partial t} + 
i \hbar \frac{2 m u_2 \dot u_2}{W} \,x\, \frac{\partial}{\partial x}
 + m^2\frac{\dot u_2^2-\omega^2 u_2^2}{W^2} \, x^2 
  + i \hbar \frac{m u_2 \dot u_2}{W} 
\label{eq:cuadraticosprimerorden1}
\\
\hat X ^2 &= \frac{\dot u_1^2-\omega^2 u_1^2}{W^2} \, x^2   
+ i \hbar \frac{2 u_1 \dot u_1}{m W} \,x\,  \frac{\partial}{\partial x}
+i \hbar \frac{2 u_1^2}{m W} \frac{\partial}{\partial t} 
+ i \hbar \frac{u_1 \dot u_1}{m W}
\label{eq:cuadraticosprimerorden2}
\\
\hat{XP} &=
-i \hbar \frac{2 u_1 u_2}{W} \frac{\partial}{\partial t} 
- i \hbar\frac{\dot u_1 u_2 + u_1 \dot u_2}{W} \,x\, \frac{\partial}{\partial x}
 - m \frac{\dot u_1 \dot u_2 -\omega^2 u_1 u_2}{W^2}\, x^2
- i  \hbar \frac{\dot u_1 u_2 + u_1 \dot u_2}{2 W}
\,,
\label{eq:cuadraticosprimerorden3}
\end{align}
which, together with $\hat X$ and $\hat P$, close the whole Schr\"odinger Lie
algebra:
\begin{align}
  & & 
  \left[ \hat X,\hat P \right] &= i \hbar & 
  &
\\
  \left[ \hat X, \hat P^2 \right] &= 2 i \hbar \hat P &  
  \left[ \hat X, \hat X^2 \right] &= 0 &
  \left[ \hat X, \hat{XP} \right] &= i \hbar \hat X
\\
  \left[ \hat P, \hat P^2 \right] &= 0 &  
  \left[ \hat P, \hat X^2 \right] &= -2 i \hbar \hat X &
  \left[ \hat P, \hat{XP} \right] &= -i \hbar \hat P 
\\ 
  \left[ \hat X^2 , \hat P^2 \right] &= 4 i \hbar \hat{XP} &  
  \left[ \hat X^2 , \hat{XP} \right] &= 2 i \hbar \hat X^2 &
  \left[ \hat P^2 , \hat{XP} \right] &= -2 i \hbar \hat P^2 
\,.
\end{align}

All these operators are well-defined on the solution space of the time-dependent
Schr\"odinger equation, so that the action of one
of them on a solution is again a solution. However, it is important to note
once again that the Hamiltonian operator corresponding to the LSODE, that is,
the quantum version of (\ref{eq:hamiltoniano general}) ($\Lambda=0$), $\hat H$,
although being a second
order differential operator, can not be expressed in terms of these operators
in general
and then it does not close a Lie algebra with them. But that which is worse, it
is not even a well-defined operator on the space of solutions of the
Schr\"odinger equation, $\mathcal H_t$.
As a consequence, $\hat H$ is not the generator of a one parameter
group corresponding to conventional time evolution.

Instead, any linear combination of $\hat P^2$, $\hat X^2$ and $\hat{XP}$,
say $\hat{\tilde H}$, can be adopted as a well-defined, Hermitian
evolution-like generator. It has an associated eigenvalue equation and real
spectrum, and its eigenvalues can be used to label solutions of
(\ref{eq:Schrodinger general}) ($\Lambda=0$) as its eigenfunctions. The
particular choice of $\hat{\tilde H}$ to be taken depends purely on convenience
and, for example, the similarity with the form of $\hat H$ of the particular
physical system. 

We would like to point out that there is an essential difference between the
approach followed in \cite{Cervero-Q} and ours. The reason is that the
un-extended Schr\"odinger group is considered there as the fundamental symmetry
of the damped harmonic oscillator, the origin of which is the analysis of the
classical equations of motion in \cite{Cervero-C}. The approach based on
QAT provides directly a representation of a central
extension of the Schr\"odinger group adapted to the specific LSODE-type
system\footnote{This fact is of the greatest relevance for the analysis of the
inclusion of time-symmetry in \cite{batenuestro} for the
damped harmonic oscillator.}. For the relevance of central
extensions in Quantum Mechanics, we refer to 
\cite{key-1,key-2,key-22,key-3,key-4,key-5}.

Let us stress that QAT can be useful to quickly perform some
calculations, avoiding tedious, direct evaluations which can become extremely
involved in the system under study. For example, it can be used to compute
the quantum propagator for any LSODE-type quantum system, following the idea of
Takagi in \cite{Takagi} for the simple case of the harmonic oscillator, or even
the evolution operator $\hat U (t)$, which becomes very difficult to evaluate
exactly when the Hamiltonian is time-dependent and does not commute with itself
at different times. 

Actually, the evolution operator of a LSODE system can be related with the free
evolution operator. Having in mind the diagram \eqref{diagram}, we write:
\begin{equation}
  \hat A \bigl( \hat U (t) \phi (x)\bigr) = \hat U_G (\tau) \varphi (\kappa)
\,.
\end{equation}
Here $\phi$ and $\varphi$ are \textit{the same} function of only one argument
($\kappa$ or $x$) and we will denote $\varphi = \phi = \psi$. Then,
\begin{multline}
  \hat U (t) \psi(x) = \hat A^{-1} \bigl(\hat U_G (\tau) \psi (\kappa) \bigr) 
= 
   \frac{1}{\sqrt{u_{2}}} e^{\frac{i}{2}\frac{m}{\hbar}
                 \frac{1}{W}\frac{\dot{u}_{2}}{u_{2}} {x}^{2}} 
    A^{*-1} \bigl(\hat U_G (\tau) \psi (\kappa) \bigr) 
= \\ =
  \frac{1}{\sqrt{u_{2}}} e^{\frac{i}{2}\frac{m}{\hbar}
                 \frac{1}{W}\frac{\dot{u}_{2}}{u_{2}} {x}^{2}} 
    A^{*-1} \bigl(\hat U_G (\tau) \bigr) A^{*-1} \bigl(\psi (\kappa) \bigr) 
\,.
\end{multline}
To factorize the function $\psi$ and single out the general action of 
$\hat U (t)$, we compute
\begin{equation}
  A^{*-1} \bigl(\psi (\kappa) \bigr) = \psi (\tfrac{x}{u_2}) = 
      e^{\log(1/u_2) x \frac{\partial}{\partial x}}\psi (x) \,,
\end{equation}
where $ e^{\log(1/u_2) x \frac{\partial}{\partial x}}$ is a dilation operator
which is \textit{not} unitary. To unitarize this operator, the generator must be
shifted from $x \frac{\partial}{\partial x}$ to $x \frac{\partial}{\partial x} +
\frac{1}{2}$, so that the true unitary operator is then 
\begin{equation}
 \hat U_D (\tfrac{1}{u_2}) = e^{\log(1/u_2) (x \frac{\partial}{\partial
x} + \frac{1}{2})} = \frac{1}{\sqrt{u_2}} e^{\log(1/u_2) x
\frac{\partial}{\partial x}} \,.
\end{equation}
But the factor $\frac{1}{\sqrt{u_2}}$ is already present in the previous
expression of $\hat U (t)$. Therefore, it now reads
\begin{multline}
  \hat U (t) =  
       e^{\frac{i}{2} \frac{m}{\hbar}\frac{1}{W}
           \frac{\dot u_2}{u_2} x^2} 
          A^{*-1} \bigl( \hat U_G (\tau) \bigr) \hat U_D (\tfrac{1}{u_2}) 
= \\ =
      \frac{1}{\sqrt{u_2}} e^{\frac{i}{2} \frac{m}{\hbar}\frac{1}{W}
           \frac{\dot u_2}{u_2} x^2} 
      e^{\frac{i \hbar}{2m} u_1 u_2 \frac{\partial^2}{\partial x^2}}
      e^{\log(1/u_2) x \frac{\partial}{\partial x}}\, .
\qquad \qquad \qquad
\label{eq:operador evolucion general}
\end{multline}
Its inverse is given by
\begin{equation} 
  \hat U (t)^{-1} = \hat U (t)^{\dagger} = \sqrt{u_2} e^{\log(u_2) x
\frac{\partial}{\partial x}}
        e^{\frac{- i \hbar}{2m} u_1 u_2 \frac{\partial^2}{\partial x^2}}
         e^{-\frac{i}{2} \frac{m}{\hbar}\frac{1}{W}
           \frac{\dot u_2}{u_2} x^2} \,.
\end{equation}

Interestingly, we have been able to obtain an \textit{exact} expression for the
evolution operator as a product of operators. No perturbative approximation
method, which could become cumbersome in some cases, is needed for \textit{any}
LSODE-related quantum system to obtain the evolution operator. These results
hold for $\Lambda \neq 0$ (see Appendix \ref{appa}).

\bigskip

As a general comment before proceeding with the computation of the wave functions, 
let us go back to the relevance of conditions
\eqref{eq:condiciones}. Those have been chosen to preserve the identity of the
variables and basic operators. 
Any other choice of solutions satisfying different initial conditions at any
given initial time would have implications which must be kept under control.
This was not taken into account in \cite{Manko}, which might result in some confusing
derivations. A general shift
\begin{equation}
  u_1 \rightarrow a u_1 + b u_2\,, \qquad
  u_2 \rightarrow c u_1 + d u_2\,,
\end{equation}
with the condition $a d - b c = 1$ 
to preserve the value of the Wronskian $W$,
is equivalent to the canonical transformation in the basic operators
\begin{equation}
  \hat X \rightarrow  a \hat X - \frac{b}{m} \hat P\,, \qquad
  \hat P \rightarrow  - c m \hat X + d \hat P \,.
\end{equation}
That is, the freedom in the choice of the solutions $u_1$ and $u_2$, which is a
$SL(2,\mathbb R)$ transformation, stands for a $Sp(1,\mathbb R)$ transformation
in basic operators and defines a family of quantum Arnold transformations. 
Then, the relation between Hilbert spaces in
the lower part of the diagram \eqref{diagram} turns into a non-trivial
transformation:
\begin{equation}
  \hat A_0 \psi(x) \equiv \hat 1 \psi (x) 
   \rightarrow
   \sqrt{d} \, e^{-\frac{i c m x^2}{2 \hbar d}} \psi (\tfrac{x}{d}) \,.
\end{equation}

\bigskip

Following the general ideas already noted, any linear combination of quadratic
operators belonging to the realization of the Schr\"odinger group (any operator in the subalgebra of
\eqref{eq:cuadraticosprimerorden1}-\eqref{eq:cuadraticosprimerorden3}) 
can be chosen in such a way that its
eigenfunctions solve \eqref{eq:Schrodinger general}. 
A specific combination of the operators
\eqref{eq:cuadraticosprimerorden1}-\eqref{eq:cuadraticosprimerorden3} 
with constant coefficients $\tilde \omega$ and $\tilde \gamma$
was already chosen in \cite{Cervero-Q}\footnote{For the damped harmonic
oscillator with constant $\omega$ and $\gamma$, coinciding with $\tilde \omega$
and $\tilde \gamma$ resp., this operator is the only one from the
$SL(2,\mathbb R)$ Schr\"odinger subalgebra which commutes with the
Hamiltonian.}:
\begin{equation}
  \hat H^{*} = \frac{1}{2m} \hat P^2 + \frac{1}{2} m \tilde \omega^2 \hat X^2 +
\frac{\tilde \gamma}{2} \hat{XP}\,. 
\label{eq:operadorg5}
\end{equation}

The eigenfunctions of this operator, solutions of the Schr\"odinger 
equation, are
\begin{multline}
 \phi_\nu(x,t) = 
\tfrac{1}{\sqrt{\sqrt{2 \pi} \Gamma(\nu+1) \sqrt{(u_2-\tilde \gamma u_1/2)^2 +
\tilde\Omega^2 u_1^2}}}
e^{
\tfrac{i}{2 \hbar} m x^2 \bigl( 
\tfrac{\tilde\Omega^2 u_1/(u_2-\tilde \gamma u_1/2)}
{(u_2-\tilde \gamma u_1/2)^2 + \tilde\Omega^2 u_1^2} 
+ \tfrac{\dot u_2-\tilde \gamma \dot u_1/2}
{(u_2-\tilde \gamma u_1/2) W}\bigr)
}
\\
\Bigl(
\tfrac{u_2-\tilde \gamma u_1/2 - i \tilde \Omega u_1}
{\sqrt{(u_2-\tilde \gamma u_1/2)^2 + \tilde\Omega^2 u_1^2}}
\Bigr)^{\nu+\frac{1}{2}}
\biggl(
C_1 D_\nu \Bigl(\tfrac{\sqrt{\tfrac{2 m \tilde\Omega}{\hbar} } x}
{\sqrt{(u_2-\tilde \gamma u_1/2)^2 + \tilde\Omega^2 u_1^2}} \Bigr) 
+ 
C_2 D_{-1-\nu} \Bigl(\tfrac{i\sqrt{\tfrac{2 m \tilde\Omega}{\hbar} } x}
{\sqrt{(u_2-\tilde \gamma u_1/2)^2 + \tilde\Omega^2 u_1^2}} \Bigr) 
\biggr)\,,
\label{eq:autoestadosg5}
\end{multline}
where $C_1$ and $C_2$ are arbitrary constants, $D_\nu$ are the parabolic 
cylinder functions \cite{Gradshteyn}, $\Gamma$ is the Gamma function,
$\tilde\Omega = \sqrt{\tilde\omega^2- \tfrac{\tilde\gamma^2}{4}}$  and $\nu$ is
in general a complex number. 

In writing $\phi_\nu (x,t)$ we have kept the generality of the quantum 
Arnold transformation so that these solutions are valid for any 
LSODE-type system (the corresponding formula for a LSODE with a external
force $\Lambda \neq 0$ is given in the Appendix \ref{appa}). This family of wave
functions is more general than the one found by Dodonov and Man'ko in
\cite{Manko} in that it contains a set of functions valid when
$\tfrac{\tilde\gamma}{2} > \tilde\omega$ even for a general LSODE
system. The associated spectrum of $\hat H^{*}$ is
\begin{equation}
 h^{*} = \hbar \,\tilde\Omega \,(\nu + \tfrac{1}{2}) \,.
 \label{espectro}
\end{equation}

To obtain these solutions, we have taken advantage of the QAT itself performing
the following steps.
We solve the time-dependent Schr\"odinger equation corresponding to a
harmonic oscillator with frequency $\tilde\Omega$, considering both the 
attractive and the repulsive cases, so that we
obtain solutions in terms of parabolic cylinder functions.  
Then, we take the QAT from this ``intermediate''
system to the free one. We compose this QAT with the inverse QAT 
corresponding to passing the free system to the present LSODE system, obtaining
this way solutions to the LSODE system Schr\"odinger equation, which are
eigenfunctions of 
$\hat H_{HO\Omega} = \tfrac{1}{2m} \hat P^2 
+ \tfrac{1}{2} m \tilde \Omega^2 \hat X^2$. 
Finally, making use of the freedom in the choice in $u_1$ and $u_2$, we perform 
the shift 
\begin{equation}
   u_1 \rightarrow u_1 \,, \qquad
   u_2 \rightarrow u_2 - \frac{\tilde \gamma}{2} u_1
  \qquad \Rightarrow \qquad 
   \tilde X \rightarrow \hat X\,, \qquad 
   \tilde P \rightarrow \hat P + m \tfrac{\tilde \gamma}{2}\hat X \,.
\end{equation}
Its effect on the quadratic operators
causes the expression of the particular combination 
\begin{equation}
  \hat H_{HO\Omega} = \tfrac{1}{2m} \hat P^2 
     + \tfrac{1}{2} m\tilde \Omega^2 \hat X^2 
\qquad \rightarrow \qquad
 \tfrac{1}{2m} \hat{P}^2 
  + \tfrac{1}{2} m \tilde \omega^2 \hat{X}^2 
  + \tfrac{\tilde \gamma}{2} \hat{XP}
= \hat H^*
\end{equation}
to change. As a consequence, the obtained solutions turn 
into $\phi_\nu(x,t)$. 

The condition of normalizability must be imposed to retain the
physically valid solutions. And we observe that the normalizability of the wave
functions depends on the specific values of $\tilde\omega$ and $\tilde\gamma$ in
the expression \eqref{eq:operadorg5}\footnote{In the specific case of the damped
harmonic oscillator, with constant $\omega$ and $\gamma$, it is possible to
identify $\tilde \omega \equiv \omega$ and $\tilde \gamma \equiv \gamma$
($\tilde \Omega \equiv \Omega$), so that the admissible wave functions vary
depending on the regime.}. For $\tilde \omega > \tfrac{\tilde \gamma}{2}$, the
normalizable solutions correspond to $C_2 = 0$ and $\nu = n$ an integer. These
functions are written then in terms of the Hermite polynomials\footnote{These
are the eigenfunctions of the operator $\hat H^*$ (there denoted as $\hat K(t)$)
found in \cite{Manko} in the general LSODE case. It should be noted that they
avoid, in this general case, the explicit mention of dimensional constants
equivalent to $\tilde
\omega$ and $\tilde \gamma$ and implicitly entrust the selection of the
specific $\hat K(t)$ to the choice of the classical solutions, which might
become rather confusing.}. In the case 
when $\tilde \Omega$ is imaginary, the solutions are Dirac-delta normalizable 
for $\nu = - \tfrac{1}{2} + i \lambda$, with $\lambda$ a real number. The 
operator $\hat H^{*}$ shows a continuous, real, doubly degenerate spectrum in 
this case \cite{AldaWolf} (see also \cite{Manko} for constant
$\omega$ and $\gamma$ in the overdamping regime). The critical case $\tilde
\Omega = 0$ can be obtained as a limit of this case.

It must be emphasized then that the choice of these
constants encodes the choice of the particular (arbitrary) quadratic operator
belonging to the Schr\"odinger algebra used to label the solutions. In the
framework of the quantum Arnold transformation, this freedom leads to
other families of solutions, different from the one presented here.

\section{Dissipative systems: Hamiltonian vs. Hermitian operators}
\label{examples}

Let us now have a close look at a couple of simple particular cases,
extensively studied in the literature: the damped particle and the damped
harmonic oscillator. Analogously, it is possible to analyze the harmonic
oscillator from the QAT point of view. Although \cite{Takagi} and \cite{Manko}
contain some aspects of this analysis, it is possible to go a bit further and
arrive at interesting results \cite{discretebasis}.

\subsection{Damped particle}
\label{exampledamped}

For $f=\gamma \,t$, $\omega=0$ and $\Lambda = 0$ in (\ref{eq:lagrangiano
general}), a Lagrangian for the damped particle can be given:
\begin{equation} 
L_{DP}=\frac{1}{2}m e^{\gamma t} \dot{x}^{2},
\label{eq:lagrangiano damped particle}
\end{equation}
where $\gamma$ is the damping constant. The equation of motion is then
\begin{equation}
\ddot{x}+\gamma\dot{x}=0.
\label{eq:ecuacion clasica particula dampada}
\end{equation}

Two independent solutions for this equation, satisfying initial conditions
(\ref{eq:condiciones}), 
\begin{equation}
 u_1(t) = \frac{1-e^{-\gamma t}}{\gamma}, \quad u_2(t)=1, 
 \quad W(t)=e^{-\gamma t}\,,
  \label{eq:us DP}
\end{equation}
provide the Arnold transformation for this system:
\begin{equation}
\begin{cases}
\tau=\frac{1-e^{-\gamma t}}{\gamma}\\
\kappa= x \\
\varphi =\phi\,,
\end{cases}
\label{eq:damped qat}
\end{equation}
which turns out to be simply a reparametrization in time. 

The Schr\"odinger equation takes the form
\begin{equation}
 i\hbar\frac{\partial \phi}{\partial t} = \hat H_{DP}\, \phi \equiv
 -\frac{\hbar^{2}}{2m}e^{-\gamma t}\frac{\partial^{2} \phi}{\partial x^{2}}\, ,
 \label{eq:Schrodinger DP}
\end{equation}
and the corresponding basic symmetry generators
\begin{equation}
 \hat P = -i \hbar \frac{\partial}{\partial x},\quad 
 \hat X = x+\frac{i \hbar}{m \gamma}(1-e^{-\gamma t})\frac{\partial}{\partial x}
 \,.
 \label{eq:operatas DP} 
\end{equation}

The crucial point is to realize that, in fact, the Hamiltonian operator
$\hat H_{DP}$ does not make sense as an operator acting on the space of
solutions of (\ref{eq:Schrodinger DP}), while $\hat P$ and $\hat X$ do. This can
be checked by direct computation. For a given solution $\phi$, the equation
satisfied by $\phi'\equiv\hat H_{DP} \phi$ is no longer (\ref{eq:Schrodinger
DP}), the reason being that $\hat H_{DP}$ does not commute with the
Schr\"odinger equation, while $\hat X \phi$, for instance, does solve it:
\begin{equation}
  \Bigl(
    i\hbar\frac{\partial}{\partial t} + 
     \frac{\hbar^{2}}{2m}e^{-\gamma t}\frac{\partial^{2}}{\partial x^{2}}
   \Bigr)
   \left( \hat X \phi \right) = 
   \Bigl(
     x+\frac{i \hbar}{m \gamma}(1-e^{-\gamma t})\frac{\partial}{\partial x}
   \Bigr)
   \Bigl(
    i\hbar\frac{\partial \phi}{\partial t} + 
     \frac{\hbar^{2}}{2m}e^{-\gamma t}\frac{\partial^{2} \phi}{\partial x^{2}}
    \Bigr) = 0 \, ,
\end{equation}
showing that the Schr\"odinger equation and $\hat X$ do commute.

There is yet another way to check explicitly that $\hat H_{DP}$ is ill-defined
in the quotient space by the time-evolution generated by itself. Formally, the
equation (\ref{eq:Schrodinger DP}) can be solved defining a
time-evolution operator $\hat U(t,t_0)$. 
The fact that $\hat H_{DP}$ commutes at different times,  
\begin{equation}
\label{eq:hamiltonianoconmuta}
 \bigl[\hat H_{DP}(t_1),\hat H_{DP}(t_2)\bigr]=0
\end{equation}
makes the calculation of $\hat U(t,t_0)$ and its action on other
operators simple: 
\begin{equation}
  \hat U(t,t_0)=e^{\frac{-i}{\hbar}\int_{t_0}^{t}\hat H_{DP}(t') \rm{d}t'} = 
  e^{\frac{i \hbar}{2 m \gamma} (e^{-\gamma t_0}-e^{-\gamma t})
     \frac{\partial^{2}}{\partial x^{2}} }\,.
\end{equation}
If we choose $t_0=0$, in agreement with conditions
(\ref{eq:condiciones}) imposed on solutions (\ref{eq:us DP}), we recover the
Arnold-transformed free evolution operator, as could be expected. However,
the computation of the evolution operator directly and the possibility of
obtaining it using QAT is not in any way trivial when
\eqref{eq:hamiltonianoconmuta} does not hold. 

By means of the action of $\hat U(t,0)\equiv \hat U(t)$ on the basic operators
$\hat P$ and $\hat X$, or loosely speaking, using $\hat U(t)$ to ``de-evolve''
them until time $t_0=0$, we can show that they match the form
(\ref{eq:operadorsubido}) and determine their action on wave functions
depending only on $x$. This action, in this simple case, can be computed by
expanding the exponential evolution operator and performing the commutation
operations at each order of the expansion, leading to 
\begin{equation}
 \hat U^{\dagger}(t)\, \hat x \,\hat U(t) = x\,,\qquad 
 \hat U^{\dagger}(t)\, \hat p \,\hat U(t) = 
    - i \hbar \frac{\partial}{\partial x}\,,
\label{eq:operatas caidos DP}
\end{equation}
which do not depend on time $t$ and take the usual Galilei form\footnote{Note
that the ``de-evolved'' operators take the form of those
in (\ref{eq:operatas DP}) when $t=0$. But in general, the correct way to take
the quotient by time evolution is that shown in (\ref{eq:operatas caidos DP}).}.
This automatically guarantees the three properties mentioned in the previous
Section. In contrast, $\hat H_{DP}$ does not come down to the quotient by the
time evolution generated by itself:
\begin{equation}
  \hat U^{\dagger}(t)\, \hat H_{DP} \,\hat U(t) =
-\frac{\hbar^{2}}{2m}e^{-\gamma t}\frac{\partial^{2}}{\partial x^{2}}\,.
\end{equation}

The reason for $\hat X$ and $\hat P$ to have good properties is that they are
mapped from the free, basic symmetry generators (\ref{eq:operata posicion
libre}) and (\ref{eq:operata momento libre}) by the Arnold transformation (so
that they are also symmetry generators of the damped particle system), while
$\hat H_{DP}$ is not. It is then natural to map one of the quadratic operators
belonging to the Schr\"odinger algebra of the free particle to make it act on
the space of quantum solutions of the damped particle. This evolution-like
operator defines a proper eigenvalue problem, that can be used to find
solutions for (\ref{eq:Schrodinger DP}). We could choose the operator 
$\hat H^*$ already mentioned,  but in this simple case we prefer to illustrate 
another rather natural possibility: the free Galilean Hamiltonian
\begin{equation}
 \hat H_G \equiv \frac{\hat P^2}{2 m} = -\frac{\hbar^2}{2 m}
\frac{\partial^2}{\partial x^2}\,.
\end{equation}
In fact, the practical approach would be to solve the free, time-independent
Schr\"odinger equation and Arnold-transform the solutions to obtain
(non-stationary) solutions for (\ref{eq:Schrodinger DP}). For example, plane
waves are mapped into:
\begin{equation}
  \phi_k (x,t) = e^{i k x - i\frac{\hbar k^2}{2 m \gamma}(1-e^{-\gamma t})} \,.
\end{equation}

The observations made above for the damped particle, being quite trivial, can
help to clarify the general case.

\subsection{Damped harmonic oscillator}
\label{exampledamposc}

Let us consider a friction function linear in time $f=\gamma t$, as in the case
of the damped particle, but also a non-zero constant frequency $\omega$ and no
external force. The Lagrangian for the Caldirola-Kanai system reads
\begin{equation} 
L_{DHO}=e^{\gamma
t}\bigl(\frac{1}{2}m\dot{x}^{2}-\frac{1}{2}m\omega^{2}x^{2}\bigr)\,.
\label{eq:lagrangianoDHO}
\end{equation}
The classical equation of motion is then
\begin{equation}
\ddot{x}+\gamma\dot{x}+\omega^{2}x=0\,.
\label{eq:ecuacionDHO}
\end{equation}

Two independent solutions of (\ref{eq:ecuacionDHO}), satisfying
convenient initial conditions (\ref{eq:condiciones}) are
\begin{equation}
u_{1}(t)=\frac{1}{\Omega}e^{-\frac{\gamma}{2}t}\sin\Omega t,\qquad
u_{2}(t)=e^{-\frac{\gamma}{2}t}\cos\Omega t + 
\frac{\gamma}{2 \Omega}e^{-\frac{\gamma}{2}t}\sin\Omega t,
\label{eq:us DHO}
\end{equation}
where again $W(t) \equiv \dot{u}_{1}(t) u_{2}(t) - u_{1}(t)\dot{u}_{2}(t) =
e^{-\gamma t}$, and                                      
\begin{equation}
\Omega=\sqrt{\omega^{2}-\frac{\gamma^{2}}{4}}\,.
\label{eq:frecuencia}
\end{equation}
Note that these solutions have good limit in the case of critical
damping $\omega = \tfrac{\gamma}{2}$.

Particularizing the quantum Arnold transformation (\ref{eq:quantum Arnold
transformation}) for the free Schr\"odinger equation (\ref{eq:Schrodinger
libre}), the Caldirola-Kanai equation is obtained:                             
\begin{equation}
i\hbar\frac{\partial \phi}{\partial t} = \hat H_{DHO}\,\phi \equiv
-\frac{\hbar^{2}}{2m} e^{-\gamma t}\frac{\partial^{2} \phi}{\partial x^{2}}
+\frac{1}{2}m\omega^{2}x^{2} e^{\gamma t}\phi \,.
\label{eq:Schrodinger DHO}
\end{equation}

Basic quantum operators are now given by
\begin{align}
 \hat P &= 
 -i \hbar \frac{e^{-\frac{\gamma t}{2}}}{2 \Omega }(2 \Omega \, \cos \Omega t
   +\gamma \, \sin \Omega t) \frac{\partial}{\partial x} 
   + m \frac{e^{\frac{\gamma t}{2}} }{4 \Omega} 
      \left(\gamma ^2+4 \Omega^2\right) \,\sin \Omega t \, x\,,
 \\
 \hat X &= 
  \frac{e^{\frac{\gamma t}{2}}}{2 \Omega } (2 \Omega \, \cos \Omega t
   -\gamma \, \sin \Omega t) \, x
    +i \hbar \frac{e^{-\frac{\gamma t}{2}}}{m \Omega }\, \sin \Omega t
     \frac{\partial}{\partial x}\,.
\end{align}
It is worth to note that these operators match those that were already found in
\cite{Manko} by hand, in looking for integrals of motion. 

Again, the key observation is that $\hat H_{DHO}$ does not make sense as an
operator acting on the space of solutions of (\ref{eq:Schrodinger DHO}), while
$\hat P$ and $\hat X$ do respect solutions. Although this can be proved by
direct calculation, it is more instructive to obtain the evolution operator
$\hat U(t,t_0)$. But the fact that $\hat H_{DHO}$ does not commute at different
times $[\hat H_{DHO}(t_1),\hat H_{DHO}(t_2)]\neq 0$ makes its calculation
trickier using conventional methods, as already mentioned. 

Generically, one would approach the problem resorting to a perturbative
method. An  appropriate method to solve the operator equation for $\hat U(t)
\equiv \hat U(t,0)$ corresponding to \eqref{eq:Schrodinger DHO} (we now make
explicit the time dependence of the Hamiltonian), 
\begin{equation}
i\hbar\frac{\partial}{\partial t} \hat U(t)= \hat H_{DHO}(t)\,\hat U(t) \,,
\label{eq:Schrodinger DHO operatorial}
\end{equation}
is that of the Magnus expansion \cite{Magnus} (see Appendix \ref{appb}). 

However, it is a good idea to take advantage of the QAT instead. The explicit
expression for the \textit{exact} evolution operator encountered for
the damped harmonic oscillator is rather involved and is found substituting
\eqref{eq:us DHO} in \eqref{eq:operador evolucion general}:
\begin{multline}
 \hat U (t) = 
\sqrt{\tfrac{2\Omega e^{\frac{\gamma}{2}t}}
{2\Omega\cos\Omega t + \gamma \sin\Omega t}} \;\;
e^{\frac{i}{2}
\frac{m}{\hbar}
\frac{-2 \omega^2 e^{\gamma t}\sin\Omega t}
{2 \Omega \cos\Omega t + \gamma \sin\Omega t}
x^2} 
\\
e^{\frac{i \hbar}{2m} 
\frac{1}{2 \Omega^2} e^{-\gamma t}
\sin\Omega t (2 \Omega \cos\Omega t + \gamma\sin\Omega t)
\frac{\partial^2}{\partial x^2}}
\;\;
e^{
\log(\frac{2 \Omega e^{\frac{\gamma}{2}t}}
{2 \Omega \cos\Omega t + \gamma\sin\Omega t}) 
x \frac{\partial}{\partial x}}\, .
\end{multline}
$\hat U^{\dagger}(t)$ is obtained analogously.
It is then possible to check the de-evolution of the operators simply expanding
the exponentials to the desired order, just as one would do for the free
evolution operator. This allows to state that:
\begin{equation}
 \hat U^{\dagger}(t)\, \hat X\, \hat U(t)=  x \,,
 \qquad
 \hat U^{\dagger}(t)\, \hat P\, \hat U(t)=  
          - i \hbar \frac{\partial}{\partial x} \,,
\end{equation}
%
% where $\mathcal O$ indicates arbitrary order.  
as expected. 

The action of $\hat U(t)$ on $\hat H_{DHO}(t)$ shows
that it does not fall down to the quotient by the time evolution 
generated by itself. In fact, the de-evolution,
\begin{multline}
 \hat U^{\dagger}(t)\, \hat H_{DHO}\, \hat U(t) = 
\\
   \Bigl(-\frac{\hbar^2}{2m}\frac{\partial^2}{\partial x^2}
       + \frac{1}{2} m \omega^2 x^2 \Bigr)
   - \gamma t \Bigl(
     -\frac{\hbar^2}{2m}\frac{\partial^2}{\partial x^2}
       - \frac{1}{2} m \omega^2 x^2 \Bigr)
\\
   + \frac{\gamma^2 t^2}{2} \Bigl(
     -\frac{\hbar^2}{2m}\frac{\partial^2}{\partial x^2}
       + \frac{1}{2} m \omega^2 x^2
     + \frac{2 \omega^2}{\gamma} \bigl(
      - i \hbar x \frac{\partial}{\partial x} - i \hbar \bigr) \Bigr)
  + \mathcal O (t^3)\,,
\end{multline}
does depend on time at any order in the time expansion.

We remark once again that there is no actual need for a
perturbative method to obtain $\hat U (t)$ in this case when QAT is used.

The computation of the wave functions, solutions of 
\eqref{eq:Schrodinger DHO}, can follow the steps shown in 
Section~\ref{quanarnold}. We select the operator $\hat H^*$ 
particularized for the damped harmonic oscillator, that is, substituting 
$u_1$ and $u_2$ by \eqref{eq:us DHO} and identifying $\tilde \omega 
\equiv \omega$ and $\tilde \gamma \equiv \gamma$. Its general 
eigenfunctions are given by the corresponding expression 
\eqref{eq:autoestadosg5}, and the spectrum \eqref{espectro}
of $\hat H^*$ will depend on the regime fixed by the specific value of 
$\Omega$.

\section{Conclusions}

The analysis carried out in this paper permits to deal with the quantum theory
of any LSODE-type dynamical system, using known properties of the quantum free
particle. The quantum Arnold transformation provides basic operators and 
establishes that the symmetry group of the free particle, the Schr\"odinger
group,
can be transferred to a realization on the LSODE system. This result turns out
to be of practical use when performing some computations, for instance
finding solutions of the Schr\"odinger equation or the evolution operator,
especially when the Hamiltonian does not commute with itself 
at different times. Even in these cases it is possible to give exact 
expressions, obtained in a non-perturbative manner. 
It is noteworthy that these calculations lead to the
knowledge of objects in the quantum theory with the only requirement
that the classical solutions of the LSODE are known.  

In a way, Arnold transformation allows to interpret LSODE-type forces,
including dissipation linear in velocity, as effects observed in a
``non-inertial reference frame''. 
The use of the present scheme goes beyond the study
of the simple damped harmonic oscillator, finding applicability in 
quite different branches of physics, such as Cosmology, where a scalar 
field appears (inflaton) satisfying equations in time which can be 
read as a LSODE. In this respect we are preparing the study of the 
specific example of the harmonic oscillator with time-dependent frequency
\cite{omegadete}. 

It is worth mentioning that the Arnold transformation used in this work is just
a particular case of a broader class of transformations which link free particle
equations to even classical non-linear equations. Research in this direction
would potentially lead to extremely useful and interesting results. We believe
that a good starting point for this purpose was presented in \cite{Mostafa2},
although some effort to establish the explicit connection with the free particle
would be in order. 

Let us end up with a final general comment concerning the symmetry 
under time translation of the quantum system associated with a LSODE. 
Even though it is possible to set up a clear framework to deal with any
LSODE-type quantum system by employing the quantum Arnold transformation, it
does not provide by itself a well-defined operator associated with proper (true)
time evolution. The reason is that conventional time 
evolution is not included in general in the symmetry group that 
can be imported from the  free system: the Hamiltonian does not belong 
to the specific representation of 
the Schr\"odinger algebra. One may wonder what happens if time evolution 
symmetry is required. We pursue this interesting issue at least 
for the damped  harmonic oscillator in \cite{batenuestro}.

\appendix
\section{Appendix: Inhomogeneous LSODE}
\label{appa}

We give here the general Arnold transformation with an extra external force
term $\Lambda$ and the corresponding generalization of the main results
above. This computation follows analogous steps as those shown before.

The general QAT is given by
\begin{equation}
\begin{cases}
\tau=\frac{u_{1}(t)}{u_{2}(t)}\\
\kappa=\frac{x-u_p(t)}{u_{2}(t)} \\
\varphi =\phi \sqrt{u_{2}(t)}\,
e^{
-\frac{i}{2}\frac{m}{\hbar}
\frac{1}{W(t)}\frac{\dot{u}_{2}(t)}{u_{2}(t)} \bigl(x-u_p(t)\bigr)^{2}
\;-\; i \frac{m}{\hbar} \frac{1}{W(t)} \dot{u}_{p}(t) x
\;-\; \frac{i}{2}\frac{m}{\hbar}
\int \frac{1}{W(t)} \bigl( u_p(t)^{2} \omega (t)^2 - \dot u_p^2 \bigr) \d t
}
\; .
\end{cases}
\label{eq:quantum Arnold transformation with external}
\end{equation}
The extra conditions to be imposed on the classical solution $u_p(t)$ to
preserve the identity of $x$ and $t$ before and after the transformation are: 
\begin{equation}
  u_p(0)=0\,, \qquad \dot u_p (0) = 0\,. 
\end{equation}
In fact, the solution $u_p(t)$ can be expressed: 
\begin{equation}
  u_p(t) = K_1(t) u_1(t) + K_2(t) u_2(t)\,,
\end{equation}
where: 
\begin{equation}
  K_1(t) = \int_0^t \frac{u_2(t')}{W(t')}\Lambda(t')\d t'\,, \qquad \qquad
  K_2(t) = -\int_0^t \frac{u_1(t')}{W(t')}\Lambda(t')\d t'\,,
\end{equation}

The transformation \eqref{eq:quantum Arnold transformation with external} leads
to the expressions for basic operators: 
\begin{equation}
\begin{split}
  \hat P &= -i \hbar u_2 \frac{\partial}{\partial x}
         - m \frac{\dot{u}_2}{W}(x-u_p) - m \frac{u_2}{W} \dot u_p \\
  \hat X &= \frac{\dot{u}_1}{W} (x-u_p) + \frac{u_1}{W} \dot u_p +
\frac{i\hbar}{m} u_1
\frac{\partial}{\partial x}\,.
\end{split}
\end{equation}
The evolution operator reads
\begin{multline}
 \hat U (t) = \frac{1}{\sqrt{u_2}} 
         e^{\frac{i}{2}\frac{m}{\hbar}
         \frac{1}{W}\frac{\dot{u}_{2}}{u_{2}} \bigl(x-u_p\bigr)^{2}
         \;+\; i \frac{m}{\hbar} \frac{1}{W} \dot{u}_{p} x
         \;+\; \frac{i}{2}\frac{m}{\hbar}
         \int \frac{1}{W} \bigl( u_p^{2} \omega (t)^2 - \dot u_p^2 \bigr)
         \d t}\\
      e^{\frac{i \hbar}{2m} u_1 u_2 \frac{\partial^2}{\partial x^2}}
      e^{-u_p \frac{\partial}{\partial x}}
      e^{\log(1/u_2) x \frac{\partial}{\partial x}}\, .
\end{multline}
Finally, the general solution of the corresponding Schr\"odinger equation, eigenfunction 
of the operator $\hat H^*$, is:
\begin{multline}
\phi_\nu(x,t) = 
\tfrac{1}{\sqrt{\sqrt{2 \pi} \Gamma(\nu+1) \sqrt{(u_2-\tilde \gamma u_1/2)^2 +
\tilde\Omega^2 u_1^2}}}
e^{ 
\bigl( 
\frac{i}{2 \hbar} m (x-u_p)^2  
\frac{\tilde\Omega^2 u_1/(u_2-\tilde \gamma u_1/2)}
{(u_2-\tilde \gamma u_1/2)^2 + \tilde\Omega^2 u_1^2} 
+ \frac{i}{2 \hbar} m x^2 \frac{\dot u_2-\tilde \gamma \dot u_1/2}
{(u_2-\tilde \gamma u_1/2) W}
\bigr)}
\\
e^{ 
\bigl( 
\frac{i m}{2\hbar}
\int \frac{1}{W} ( u_p^{2} \omega (t)^2 - \dot u_p^2 ) \d t
- \frac{i m x u_p (\dot u_2-\tilde \gamma \dot u_1/2)}
{\hbar W(u_2-\tilde \gamma u_1/2)}
+ \frac{i m u_p^2 (\dot u_2-\tilde \gamma \dot u_1/2)}
{2 \hbar W(u_2-\tilde \gamma u_1/2)}
+ \frac{i m x \dot u_p}{\hbar W}
\bigr)
}
\\
\Bigl(
\tfrac{u_2-\tilde \gamma u_1/2 - i \tilde \Omega u_1}
{\sqrt{(u_2-\tilde \gamma u_1/2)^2 + \tilde\Omega^2 u_1^2}}
\Bigr)^{\nu+\frac{1}{2}}
\biggl(
C_1 D_\nu \Bigl(\tfrac{\sqrt{\tfrac{2 m \tilde\Omega}{\hbar} } (x-u_p)}
{\sqrt{(u_2-\tilde \gamma u_1/2)^2 + \tilde\Omega^2 u_1^2}} \Bigr) 
+ 
C_2 D_{-1-\nu} \Bigl(\tfrac{i\sqrt{\tfrac{2 m \tilde\Omega}{\hbar} } (x-u_p)}
{\sqrt{(u_2-\tilde \gamma u_1/2)^2 + \tilde\Omega^2 u_1^2}} \Bigr) 
\biggr)\,.
\end{multline}

\section{Appendix: The Magnus expansion}
\label{appb}

The Magnus expansion was introduced as a tool to
solve non-autonomous linear differential equations for linear operators and has
the very attractive property of leading to approximate solutions which exhibit
unitarity at any order of approximation. This is in contrast to
the representation in terms of the time-ordering operator
$\mathcal{T}$ introduced by Dyson.

A solution to \eqref{eq:Schrodinger DHO operatorial} is given by
\begin{equation}
  \hat U(t) = e^{\hat \Omega (t)}\,, \qquad \hat \Omega(0) = 0 \, ,
\end{equation}
and a series expansion for the matrix in the exponent
\begin{equation}
  \hat \Omega (t) = \sum_{k=1}^{\infty} \hat \Omega_k (t)
\end{equation}
which is called the Magnus expansion. We can write down the first three
terms of that series: 
\begin{align}
  \hat \Omega_1(t) &= 
          \int_0^t \d t_1 \Bigl(-\frac{i}{\hbar}\hat H_{DHO}(t_1) \Bigr) 
\\
  \hat \Omega_2(t) &=\frac{1}{2}\int_0^t \d t_1 \int_0^{t_1} \d t_2
     \Bigl[ 
       -\frac{i}{\hbar}\hat H_{DHO}(t_1),\, -\frac{i}{\hbar}\hat H_{DHO}(t_2)
     \Bigr]
\\
   \begin{split}
  \hat \Omega_3(t) &=
      \frac{1}{6}\int_0^t \d t_1 \int_0^{t_1} \d t_2 \int_0^{t_2} \d t_3
     \Biggl(
       \biggl[ -\frac{i}{\hbar}\hat H_{DHO}(t_1)\,,
         \Bigl[ 
            -\frac{i}{\hbar}\hat H_{DHO}(t_2)\,,
            -\frac{i}{\hbar}\hat H_{DHO}(t_3)
         \Bigr]
       \biggr]
       \\  & \qquad \qquad \qquad  +
       \biggl[ -\frac{i}{\hbar}\hat H_{DHO}(t_3)\,,
         \Bigl[
           -\frac{i}{\hbar}\hat H_{DHO}(t_2) \,, 
           -\frac{i}{\hbar}\hat H_{DHO}(t_1)
         \Bigr]
        \biggr]
      \Biggr)
     \end{split}
\,.
\end{align}
%
% 
% Then, if all terms in ME have 
% a structure similar to that of the ones shown before, the whole $\Omega(t)$
% and
% any approximation to it obtained by truncation of 
% ME, will also belong to the same Lie algebra. This ensures unitarity: every 
% approximant preserves probability conservation.
% 
% Existence of a solution, convergence of the series, general expression for the
% series, efficient computation. 

However, a good iterative method to obtain the operator $\hat \Omega (t)$ is
given by the formula:
\begin{align}
  \hat \Omega^{[n]} (t) &= \sum_{k=0}^{\infty}\frac{B_k}{k!}\int^{t}_{0} \d t_1
   \rm{ad}^k_{\hat \Omega^{[n-1]}(t_1)}
       \bigl(-\frac{i}{\hbar}\hat H_{DHO}(t_1)\bigr)
\label{eq:Omegan}
\\
   \hat \Omega (t) &= \lim_{n\to\infty}\hat \Omega^{[n]}(t)\,,
\end{align}
where $B_k$ are the Bernoulli numbers and 
\begin{equation}
\rm{ad}^0_{\hat A} (\hat B)\equiv \hat B\,, \qquad
\rm{ad}^1_{\hat A} (\hat B)\equiv [\hat A, \hat B]\,, \qquad
\rm{ad}^k_{\hat A}(\hat B)\equiv [\rm{ad}^{k-1}_{\hat A}(\hat B),\hat B]\,.
\end{equation}

We have computed the operator $\hat \Omega(t)$ for the case of the damped
harmonic oscillator to sixth order of approximation, to give: 
\begin{multline}
  \hat \Omega^{[6]} (t)  =
\\ 
-\frac{i}{\hbar}t
   \biggl(
    \Bigl(
      1 + \tfrac{\gamma^2 t^2}{6} 
        + \tfrac{\gamma^4 t^4}{120} (1+\tfrac{2 \omega^2}{\gamma^2})
        + \tfrac{\gamma^6 t^6}{5040} 
          (1 + \tfrac{16 \omega^2}{\gamma^2} + \tfrac{32 \omega^4}{3 \gamma^4})
    \Bigr) 
      \Bigl(
       -\frac{\hbar^2}{2m}\frac{\partial^2}{\partial x^2}
       + \frac{1}{2} m \omega^2 x^2
      \Bigr) 
 \\ \qquad
  - \frac{\gamma t}{2}
    \Bigl(
      1 + \tfrac{\gamma^2 t^2}{12}
        + \tfrac{\gamma^4 t^4}{360} (1+\tfrac{6 \omega^2}{\gamma^2})
    \Bigr)
      \Bigl(
       -\frac{\hbar^2}{2m}\frac{\partial^2}{\partial x^2}
       - \frac{1}{2} m \omega^2 x^2
      \Bigr) 
 \\
  + \frac{\gamma \omega^2 t}{6}
    \Bigl(
      1 + \tfrac{\gamma^2 t^2}{20} (1 + \tfrac{4 \omega^2}{3 \gamma^2})
        + \tfrac{\gamma^4 t^4}{840} 
            (1 + \tfrac{44 \omega^2}{3 \gamma^2}
                + \tfrac{16 \omega^4}{3\gamma^4})
    \Bigr)
      \Bigl(
       - i \hbar x \frac{\partial}{\partial x}
       - \frac{1}{2} i \hbar
      \Bigr)
   \biggr) 
\end{multline}
With this approximation to $\hat \Omega (t)$, obtained by means of
\eqref{eq:Omegan}, one can obtain the quotient by time evolution of a certain
operator $\hat O(t)$, given by:
\begin{equation}
  \hat O = e^{-\hat \Omega (t)} \hat O (t) e^{\hat \Omega (t)} = 
    \sum_{k=0}^{\infty} \frac{1}{k!} 
     \rm{ad}^k_{-\hat \Omega (t)} \bigl( \hat O (t) \bigr)\,.
\end{equation}

These formulas lead to the same results found in Subsection
\ref{exampledamposc}.
% 
% On the contrary, the de-evolution of $\hat X$ and $\hat P$ leads to the same
% results as in the damped particle (\ref{eq:operatas caidos DP}), as expected.
% To this order of approximation, the expressions we have encountered for them
% are the usual ones: 
% %
% \begin{equation}
%  \hat U^{\dagger}(t)\, \hat X\, \hat U(t)=  x + \mathcal O (t^7)\,,
%  \qquad
%  \hat U^{\dagger}(t)\, \hat P\, \hat U(t)=  
%           - i \hbar \frac{\partial}{\partial x} + \mathcal O (t^7) \,.
% \end{equation}
%

\section*{Acknowledgments}

Work partially supported by the Fundaci\'on S\'eneca, Spanish MICINN and
Junta de Andaluc\'\i a under projects 08814/PI/08, FIS2008-06078-C03-01 
and FQM219-FQM1951, respectively. 

The authors wish to thank M. Calixto for useful discussions and comments.

\end{document}